\documentclass[11pt]{cernrep}

\begin{document}

\begin{flushright}
MAN/HEP/2006/25 \\
20 September 2006 \\

\end{flushright}
\begin{center}
{\par \noindent \textbf {\large The FP420 R\&D Project at the LHC}\large \par}
\bigskip{}
{\par \centering B. E. Cox} \\
{\par \noindent School of Physics and Astronomy, University of Manchester, \\
Manchester M13 9PL, UK \\
}
\end{center}

\section{Introduction and Motivation for FP420}

FP420 is an R\&D collaboration formed to investigate the 
feasibility of installing forward proton tagging detectors in a 15m-long region 420m from the interaction points of ATLAS and / or CMS. These detectors are envisaged to be sub-detector upgrades to the central detectors, which could be installed at a suitable time after the initial phase of LHC running. The aim is to make use of the unique properties of 
the central exclusive production process to extend the capabilities of ATLAS and CMS to 
search for and identify the nature of new particles.     
By central exclusive, we refer to the process $PP\rightarrow P \oplus \phi \oplus P$, where 
$\oplus$ denotes the absence of hadronic activity ('gap') between the outgoing protons and the 
decay products of the central system $\phi$. An example would be standard model Higgs Boson production, 
where the central system may consist of 2 $b$-quark jets, and {\it no other activity}. For the 
production of central systems in the 120 GeV mass range, the outgoing protons emerge from the beam envelope in the high-dispersion 420m region when the LHC runs with standard high-luminosity optics.   
The process is attractive for two main reasons. Firstly, if the outgoing protons remain intact and scatter
 through small angles, then, to a very good approximation, the central system $\phi$ must be produced in a 
spin $0$, CP even state, therefore allowing a clean determination of the quantum numbers of any observed resonance. Secondly, using the so-called missing mass method \cite{Albrow:2000na}, a resolution on the central system of the order of 1 GeV is achievable by measuring the momentum loss of the outgoing protons alone, irrespective of the decay mode of the produced central particle \cite{Alekhin:2005dy}. The Standard Model Higgs Boson is predicted to be observable in the central exclusive channel with $30$ fb$^{-1}$ of delivered luminosity \cite{DeRoeck:2002hk,Cox:2005if}.
The simplest decay channel from an experimental perspective is the $WW / WW^*$ decay mode, in which 
one (or both) of the W Bosons decays leptonically. With standard single
 and double lepton trigger thresholds at ATLAS and CMS, approximately 5 events are 
expected with double proton tags for Higgs Boson masses in the range $140$ GeV $< M_H <200$ GeV 
with 30 fb$^{-1}$ of LHC luminosity (10 with slightly reduced trigger thresholds) \cite{Cox:2005if}. The $b \bar b$ channel should also be observable because the quantum number selection rules in central exclusive production strongly 
suppress the QCD $b$-jet background, and the central exclusive cross sections are certainly 
expected to be large enough \cite{Khoze:2001xm}. Unlike the $WW$ decay 
mode, however, triggering at level 1 at ATLAS and CMS presents a potential difficulty, 
because the 420m 
detectors are too far away from the central detectors to be included in the level 1 triggers of 
either experiment in normal running mode. In order to access the $b \bar b$ decay channel, 
therefore, a means must be found of triggering on the central detectors alone. In the case of no pile-up, it is
possible to lower the level 1 jet trigger thresholds by a sufficient margin to retain the Higgs 
events by vetoing on energy in the forward region at level 1 \cite{Alekhin:2005dy}.
At luminosities up to $2 \times 10^{33}$cm$^{-2}$s$^{-1}$ it is possible to save events in which 
one proton is detected in a 220m detector, (the TOTEM detectors at CMS or upgraded luminosity roman pots at ATLAS), which can be triggered on at level 1, although such asymmetric double tagged events have a poorer 
mass resolution \cite{Alekhin:2005dy}. A further 10\% of $b$-jet events can be saved irrespective of pile-up conditions by triggering on the muons from $b$-quark decays in the jets \cite{Alekhin:2005dy}. In certain regions of the MSSM parameter space the 
cross section for the central exclusive production of the lightest Higgs Boson is significantly 
enhanced, and in these scenarios the muon triggers alone will deliver sufficient events that 
double proton tagging may be the discovery channel \cite{Kaidalov:2003ys}. Another attractive 
feature is the ability to directly probe the CP structure of the  Higgs sector by measuring 
azimuthal asymmetries in the tagged protons (a measurement previously proposed only at a future 
linear collider) \cite{Khoze:2004rc}.

The event yields given above depend of course on the accuracy of the theoretical calculations of 
central exclusive production (a recent review can be found in \cite{Forshaw:2005qp}). The CDF 
Collaboration have carried out a search for an exclusive component in Double 
Pomeron Exchange (DPE) events at the Tevatron \cite{Gallinaro:2006vz,Royon:2006qg}. If the Khoze, Martin and Ryskin (KMR) theoretical framework \cite{Khoze:2001xm} used for the LHC predictions is correct, it was shown in \cite{Cox:2005gr} that an exclusive component should be visible in the Tevatron 
DPE data, and indeed this appears to be the case although there are uncertainties due to 
the uncertainty in the knowledge of the high $\beta$ gluon distribution in the pomeron. Perhaps 
more significant, however, is the CDF measurement of the heavy flavour fraction in the DPE di-jet sample \cite{Gallinaro:2006vz}. It is a prediction of KMR that the 
$gg \rightarrow q \bar q$ channel is suppressed relative to the $gg \rightarrow gg$ channel in central exclusive production. The CDF collaboration do see a significant reduction 
in the fraction of tagged c and $b$-quark jets in the region where the exclusive component is expected. CDF also 
searched for exclusive di-photon events, which again should be visible in the Tevatron data according to the KMR calculations (although it should be stressed that there are significant theoretical uncertainties due to the small $\gamma \gamma$ invariant masses). CDF observe 3 exclusive di-photon events which is compatible with KMR \cite{Gallinaro:2006vz}. 
\section{Status of the FP420 R\&D project}

At LHC start-up, the beam pipes in the 420m region are contained within a 15m-long interconnection cryostat that connects the superconducting arcs and dispersion suppressor regions of the LHC. The 
cryostat provides continuity not only of the 2 K cold beam-pipes, but also of the insulation vacuum, electrical power, 
cryogenic circuits and thermal and radiation shielding. The initial challenge for FP420 was to design a new 
15m section to replace the interconnection cryostat and allow access to warm beam pipes. This has 
been achieved using to a large extent existing LHC components, reducing the 
need for extensive prototyping, testing and re-tooling. Full engineering design work is now underway.   
Design and prototyping of the FP420 detectors and mechanical and electrical systems is also proceeding well. Detection of the proton tracks will be achieved by two 3D silicon detector stations at each end of the FP420 region. A position accuracy of $~50$ microns is required to deliver the 
maximum theoretical energy resolution of $~ 1$ GeV, which is set by the uncertainty in the LHC beam energy. It is highly desirable to aim to achieve position accuracy at the sub-10 micron level in
order to measure the angle of the proton tracks through the FP420 region, however, since measurements of the 
proton transverse momenta gives access to a wealth of physics including direct observation of CP violation in 
the Higgs sector. The baseline FP420 design also incorporates both QUARTZ and Gas Cherenkov fast-timing 
detectors capable of sub-10 ps resolution, allowing for a determination of the vertex position of 
double-tagged proton events to within $2.1$ mm. Correlating this position with a 
central vertex position allows for rejection of background proton tracks with an efficiency of  97\%, which is 
sufficient to operate at luminosities of $2 \times 10^{33}$cm$^{-2}$s$^{-1}$. Further details of the base-line FP420 concept can be found in \cite{FP420LHCC}.            
\section{Summary}
There is mounting evidence in several different physics channels that there is an exclusive component present in the CDF data, and that the overall normalisation and event characteristics are consistent with the KMR predictions. This suggests that the central exclusive production channel could be exploited at the LHC to extend the discovery reach if suitable detector upgrades are installed. The FP420 collaboration expects to have a detailed feasibility study and design to present to the ATLAS and CMS collaborations during the first half of 2007.         \section*{Acknowledgments}
We thank the AT-CRI, TS-MME and TS-LEA groups at CERN for their continuing help and support in the FP420 design study. The author is supported by The Royal Society and PPARC in the UK.

\end{document}